\magnification = \magstep1
\pageno=0
\hsize=15.0truecm
\hoffset=3.0truecm
\hoffset=0.5truecm
\vsize=22.5truecm
\voffset=0.0truecm

\output={\plainoutput}
\pretolerance=3000
\tolerance=5000
\hyphenpenalty=10000    
\newdimen\digitwidth
\setbox0=\hbox{\rm0}
\digitwidth=\wd0

\def\footnoterule{\kern-3pt \hrule width \hsize \kern 2.6pt
\vskip 3pt}
\def\cl{\centerline}
\def\ni{\noindent}

\def\vs{\vskip 11pt}

\def\solar{\ifmmode _{\mathord\odot}\else $_{\mathord\odot}$\fi}

\font\ksub=cmsy7
\def\teff{T$_{\kern-0.8pt{\ksub e\kern-1.5pt f\kern-2.8pt f}}$}

%
\vs\vs\vs
\vs\vs\vs
\vs
\vs
\vs
\vs
\vs
\cl{\bf MOST POPULATION III SUPERNOVAE ARE DUDS}
\vs\vs
\cl{Robert L. Kurucz}
\cl{Harvard-Smithsonian Center for Astrophysics}
\vs
\vs
\cl{August 22, 2008}
\vs
\vs\vs\vs 
\eject
\vs\vs
\cl{\bf MOST POPULATION III SUPERNOVAE ARE DUDS}
\vs\vs
\cl{Robert L. Kurucz}
\vs
\cl{Harvard-Smithsonian Center for Astrophysics}
\cl{60 Garden St, Cambridge, MA 02138}
\vs
\centerline {\bf Abstract}

     One Population III dud supernova produces enough oxygen to 
enable 10$^{7}$ solar masses of primordial gas to bind into M dwarfs.  
This is possible because radiation from other Population III stars 
implodes the mixture of oxygen ejecta and primordial gas into a 
globular cluster.  Model atmosphere calculations for oxygen dwarfs 
show that water blocks most of the infrared flux.  The flux is 
redistributed into the visible to produce an unfamiliar, distinctive 
energy distribution.  One million dud supernovae in a large 
protogalaxy are sufficient to produce the ``dark matter" halo.
\vs
Subject headings: supernovae --- dark matter
\vs
\vs
\centerline {\bf Introduction}
\vs

Most of the physics and literature on Population III star formation
and on Population III supernovae are 
speculative, as is this paper.  The goal of supernova modelling
is always to produce supernova explosions and products.  Methods and 
physics that do not work to produce an explosion on the computer 
are not interesting, are not funded, and do not produce jobs.  
How can we ever discover how to make duds, or that duds even exist?
Here I make predictions about dud supernovae and about ``dark matter".
\vs
\vs     
\centerline{\bf Population III stars, dud supernovae, supernovae, oxygen dwarfs}
\vs
     In prestellar times small perturbations cool by radiating in 
HD, LiH, and other light molecules and start to collapse.  If the 
perturbations are too massive or have too much angular momentum, 
they cannot successfully collapse to form a star.  

     At some threshold of mass or angular momentum, a small perturbation 
($>$ 1000 M$_\odot$) can collapse by somehow ejecting enough mass and 
angular momentum to produce a Population III star of, say, 200 M$_{\odot}$ 
that is rapidly rotating and oblate.  As it evolves, the star continues to 
lose a significant fraction of its mass and angular momentum through a 
radiatively-driven wind from the loosely bound mass at the equator.  
The star produces a huge 
prolate ``Stromgren sphere".  The core of the star is rapidly 
rotating, oblate, and hotter and faster burning at the poles and 
along the axis than in the equatorial plane.  There is a strong 
meridional circulation that mixes in fresh fuel.  The star radiates 
about 10$^{51}$ ergs/M$_{\odot}$ 
over its lifetime that is only on the order of 10$^{5}$ years.  It 
burns to an oxygen core, but then, because of the rapid rotation, the 
oblateness, and the rapid evolution, the star cannot stably collapse.
It flies apart in the attempt.  It explodes with, say, 10$^{-3}$ of 
the energy of a supernova and produces ejecta that move at only 
10$^{3}$ km s$^{-1}$ instead of 3$\times$10$^{4}$ km s$^{-1}$.  It is a dud that produces, say, 
25 M$_{\odot}$ of oxygen, unburned dregs of lighter elements, no heavier 
elements, and no remnant.  The ejecta mix with the previously lost 
mass and with the surrounding primordial gas, but much more slowly 
than would a supernova remnant.  

     Larger perturbations produce (several) smaller Population III stars with 
less angular momentum. One of these stars, with 150 to 100 M$_{\odot}$, 
evolves more slowly and stably than a more massive star.  It is less 
rapidly rotating and less oblate.  It loses a significant fraction 
of its mass and angular momentum through a radiatively-driven wind 
at the equator.  It produces a prolate ``Stromgren sphere".  The star 
radiates about 10$^{51}$ ergs/M$_{\odot}$ over its lifetime that is on the 
order of 10$^{6}$ to 10$^{7}$ years.  It may also be a dud, but it likely 
produces a supernova with iron or alpha elements, or both, and leaves 
behind a black hole.

     When radiation from a Population III star hits an oxygen blob 
from a dud supernova, it compresses the gas and the gas radiates in 
OH and H$_{2}$O to cool rapidly.   Ultraviolet radiation from the star is 
down-converted to visible-, infrared-, and radio-line radiation that 
cannot be absorbed by the primordial gas surrounding the oxygen blob 
but is strongly absorbed within the blob (and by other blobs).  Any 
other elements present from the dud explosion add to the molecular 
mix and increase the line opacity.  The illuminated blob surface 
forms a wall of low mass oxygen-dwarf stars.  The process repeats 
forming layers of stars.  Radiatively-driven implosions are described 
in Kurucz (2000).  Stars formed in this way have not yet evolved off 
their main sequence at the present time.  

     I have computed ATLAS12 (Kurucz 1995; 2005) model atmospheres 
for oxygen dwarfs with Teff 3500K and 3000K; log g = 5; for fractional 
abundances by number H = .911, He = .089, log Li = -10, log O = -4,-5,-6.  
The pinch of Li adds a few electrons.  Figure 1 shows the temperature-Rosseland
optical depth relations compared to those for Population I and
extreme Population II M dwarfs of the same Teff and gravity.  Note 
that these are photospheric models with no temperature minimum 
and no chromosphere.  The temperature distributions are radically 
different from those of Population I and Population II stars.  Figures 2 
through 7 show the energy distributions for these models.  Strong water 
absorption in the infrared of the oxygen dwarfs forces the stars to 
radiate in the visible to maintain energy 
balance.  Oxygen dwarfs have a nearly featureless spectrum in the
visible except for H$\alpha$ which would be affected by a chromosphere.  
If the parent dud supernova had some carbon, the oxygen dwarf daughters 
bind the carbon in CO which produces additional absorption in the infrared.

     Let there be, say, 10$^{6}$ dud supernovae per large protogalaxy.
The protogalaxies become full of oxygen blobs.  Let there be, say,
10$^{6}$ Population III stars smaller than 150 M$_{\odot}$ that radiate
10$^{51}$ ergs/M$_{\odot}$ and then supernova.  If the Population III stars
and the oxygen blobs are randomly positioned relative to each other
in the protogalaxy, the blobs will be illuminated on all sides by 
radiation from the Population III stars.  Four Population III stars
arranged in a tetrahedron are sufficient to implode an oxygen 
blob of 10$^{7}$ M$_{\odot}$ into an oxygen-dwarf globular cluster.  One 
million globular clusters become the ``dark matter" halo of large
galaxies.  

     As the oxygen blobs are used up by forming stars, they are 
replaced by the Population III supernova remnants that have iron or 
alpha elements or both.  Supernova remnants are energetic and mix 
through much larger volumes of primordial gas than the oxygen 
blobs were able to; the average abundances are much lower.  The 
admixture of metals to the primordial gas increases its opacity 
and makes it possible to form Population II stars much smaller 
than Population III stars.  Radiatively-driven implosions produced 
by less massive, longer-lived Population III stars yield individual Population 
II stars and Population II globular clusters.  Massive Population II
stars also produce implosions. The Population II mass function 
varies from high to a low limit set by the local opacity. 

     The galaxy-size perturbations that fill the universe are 
themselves filled with oxygen-dwarf and Population II globular 
clusters, with oxygen-dwarf and Population II field stars, and 
with black holes from Population III supernovae.  Most of the gas 
has been formed into stars.  The evolutionary details are discussed 
in my paper on radiatively-driven cosmology (Kurucz 2000).  Without 
much gas, stars and clusters act dynamically like point masses.  
The globular clusters violently relax into elliptical galaxies.  
The globular clusters eventually lose stars or accrete stars or 
disintegrate or merge.  Only a few peculiar Population II globular 
clusters that are dominated by low mass stars, and a few dwarf 
galaxies, are still visible, although there can be many that are 
too faint to see.  ``Spiral galaxies" are just minor structures 
made from Population II supernova remnants and Population II mass 
loss material that collect at and around the centers of large, 
mostly invisible, elliptical galaxies.
\vs
\centerline{\bf Prediction}
\vs
     The ``dark matter" halo consists of 10$^{6}$, or so, oxygen-dwarf 
globular clusters, up to 25 per square degree.  The clusters would appear
10" to 60" in diameter and would have thousands of stars per square 
arcsecond, but would still be transparent.  The observational test is to 
produce [M$_{RED}$ -- M$_{IR}$] color maps in various filter 
systems with the galaxies and stars blocked out.  The globular clusters
would appear as bumps of color excess.  The ``dark 
matter" halo also has 10$^{12}$ to 10$^{13}$ oxygen dwarfs in the field 
that have escaped from clusters.  There could be up to 20 oxygen dwarfs
per square arcsecond that would not be individually detectable.
\vfill
\eject
\centerline{\bf References}

\ni Kurucz, R.L. 1995. A new opacity-sampling model atmosphere program for 

arbitrary abundances.  in Stellar Surface Structure, ed. K.G. Strassmeier and 

J.L. Linsky (Dordrecht: Kluwer), 523-526.

\ni Kurucz, R.L. 2000.  An outline of radiatively driven cosmology. 

(astro-ph/0003381).

\ni Kurucz, R.L. 2005.  Memorie della Societa Astronomica Italiana Supplementi 8, 

10-20.
\vs
\vs
\vs
\centerline{Figure captions}
\vs
\ni Figure 1. Temperature-log Rosseland optical depth relations for model atmo- 
spheres with Teff 3500K, log g 5.0 .  The colors are the same in all
the figures.  Red is Population I solar abundance.  Green is Population II 
1/300 solar abundance with alpha elements enhanced by 2.5 .  Cyan is an oxygen 
dwarf with log fractional O abundance by number = -4.  Purple is an oxygen 
dwarf with log fractional O abundance by number = -5.  Black is an oxygen 
dwarf with log fractional O abundance by number = -6.  
\vs
\ni Figure 2. Energy distributions at resolving power 500 for the wavelength
range 0 to 3000 nm.  The smooth curves represent the continuum level.
The models shown are Population I, Population II, and oxygen dwarf -4.
Full scale is 1.5$\times$10$^{7}$ ergs cm$^{-2}$ s$^{-1}$ nm$^{-1}$ at
the star.
\vs
\ni Figure 3. As Figure 2 but for the wavelength range 3000 to 10000 nm.
Full scale is 7.5$\times$10$^{5}$ ergs cm$^{-2}$ s$^{-1}$ nm$^{-1}$
at the star.
\vs
\ni Figure 4. As Figure 2 but for oxygen dwarfs -4, -5, and -6.
\vs
\ni Figure 5. As Figure 4 but for the wavelength range 3000 to 10000 nm.
\vs
\ni Figure 6. As Figure 4 but for Teff 3000K, log g 5.0 .
\vs
\ni Figure 7. As Figure 6 but for the wavelength range 3000 to 10000 nm.

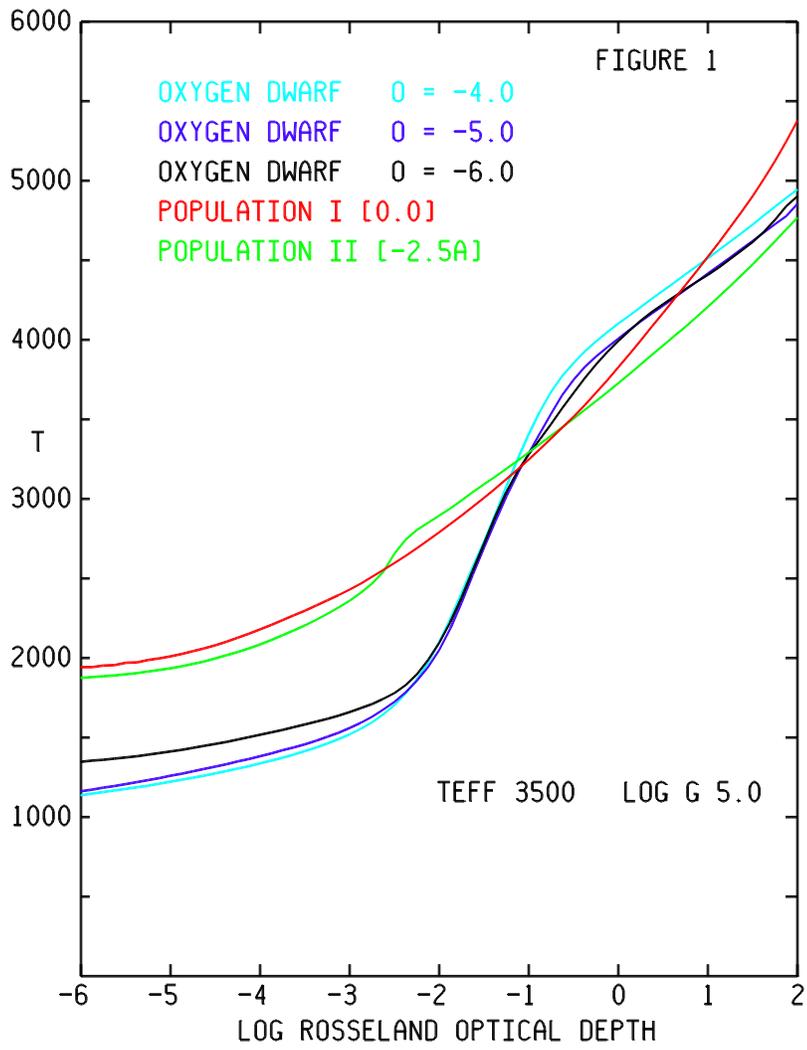
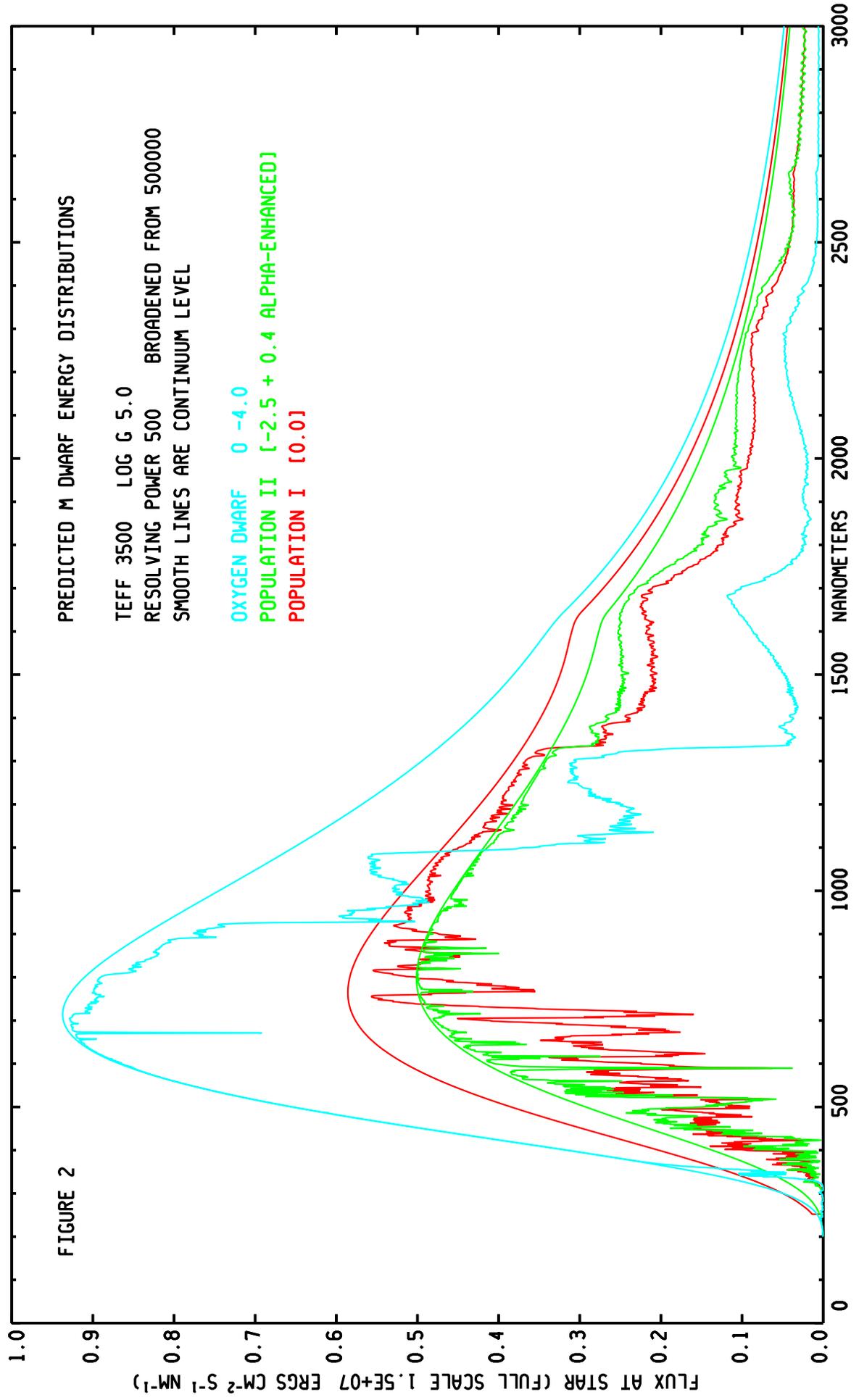
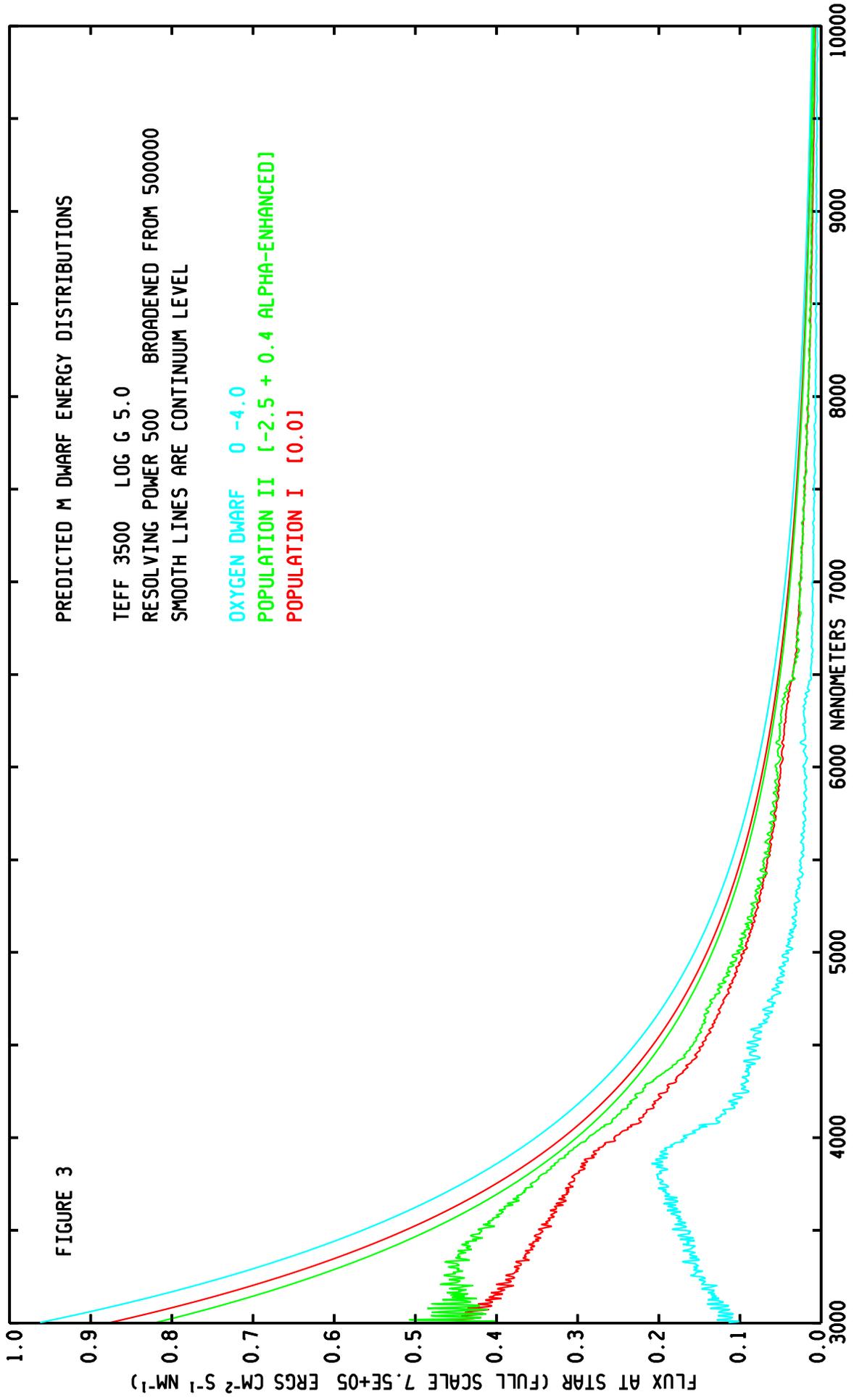
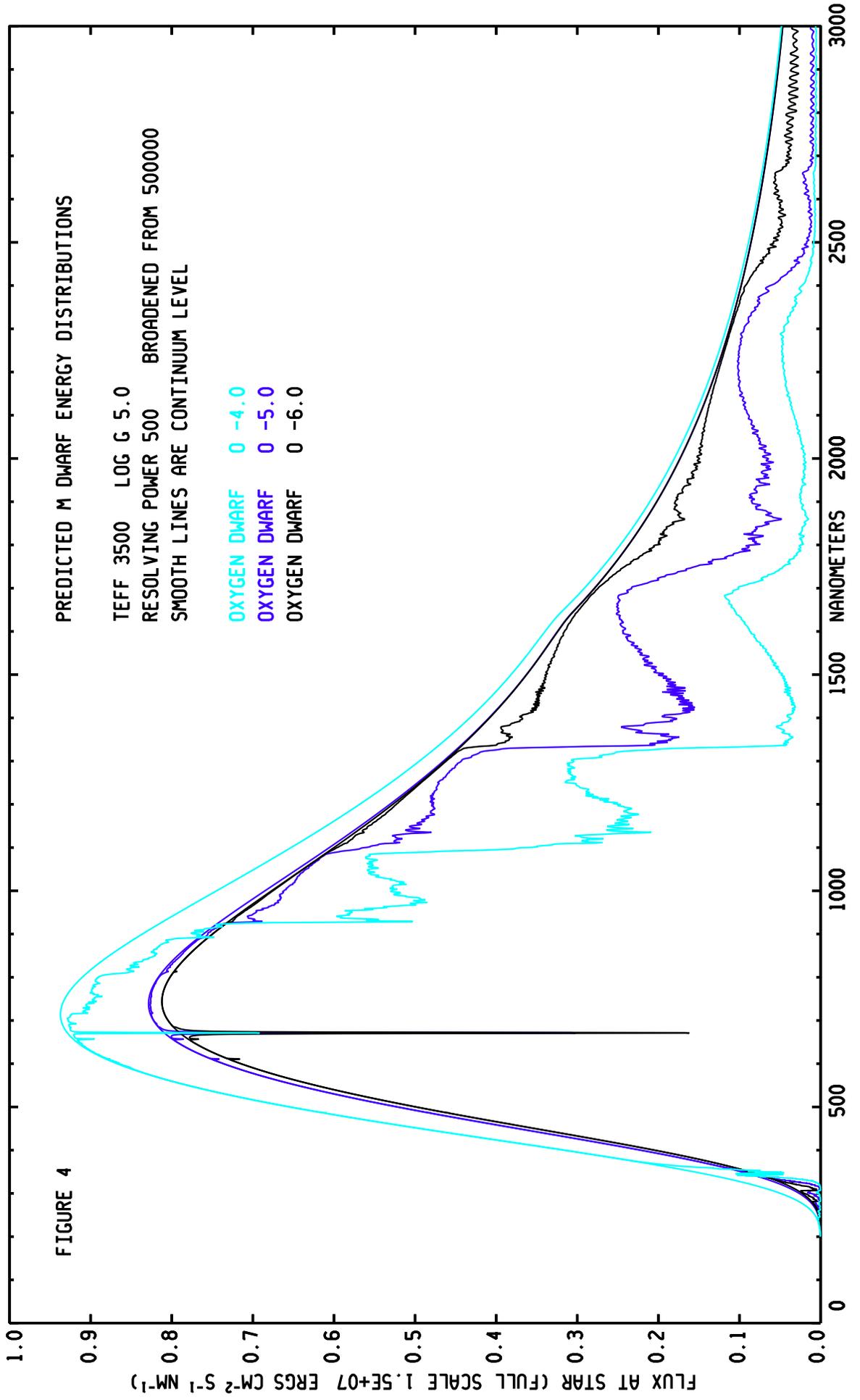
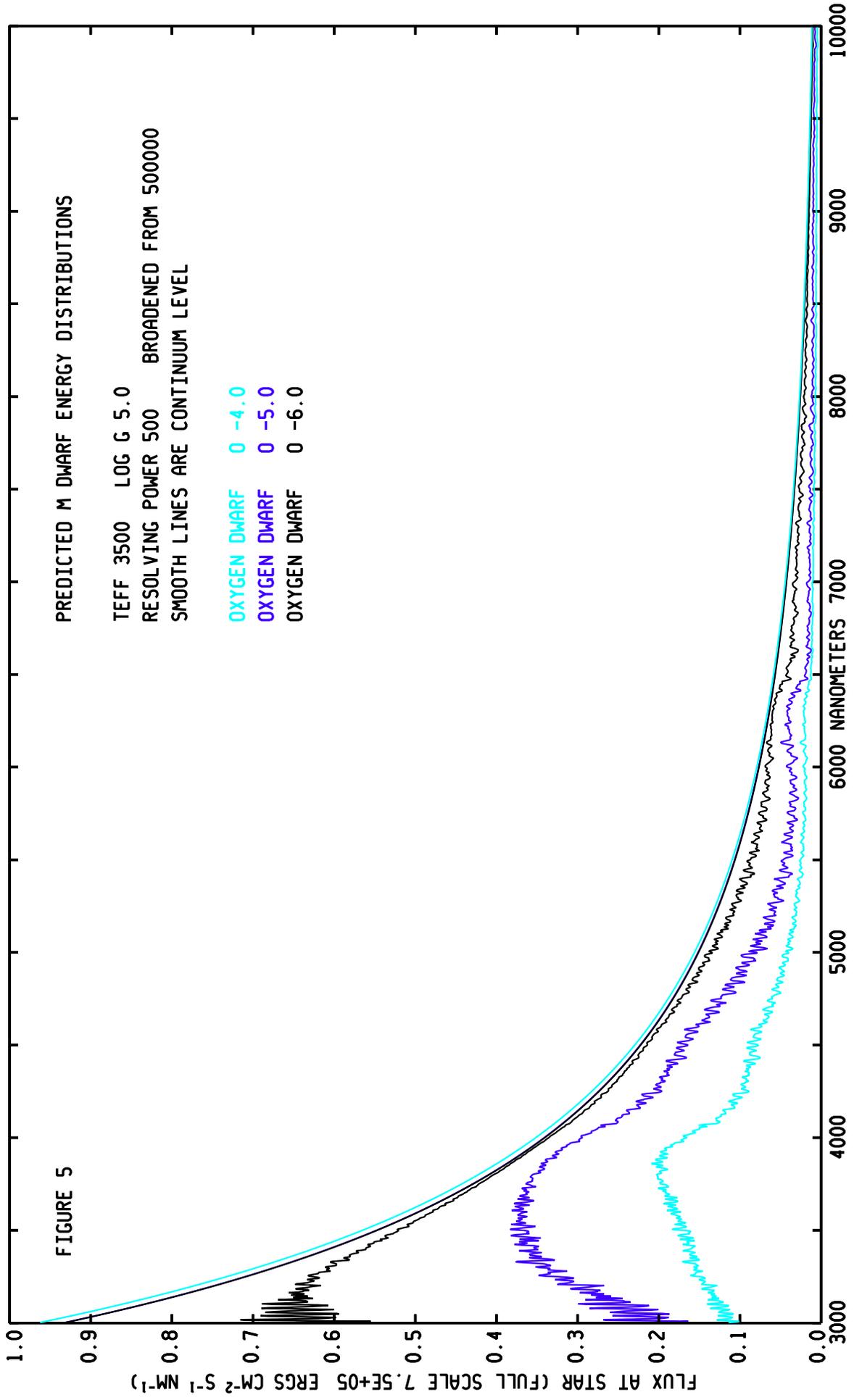
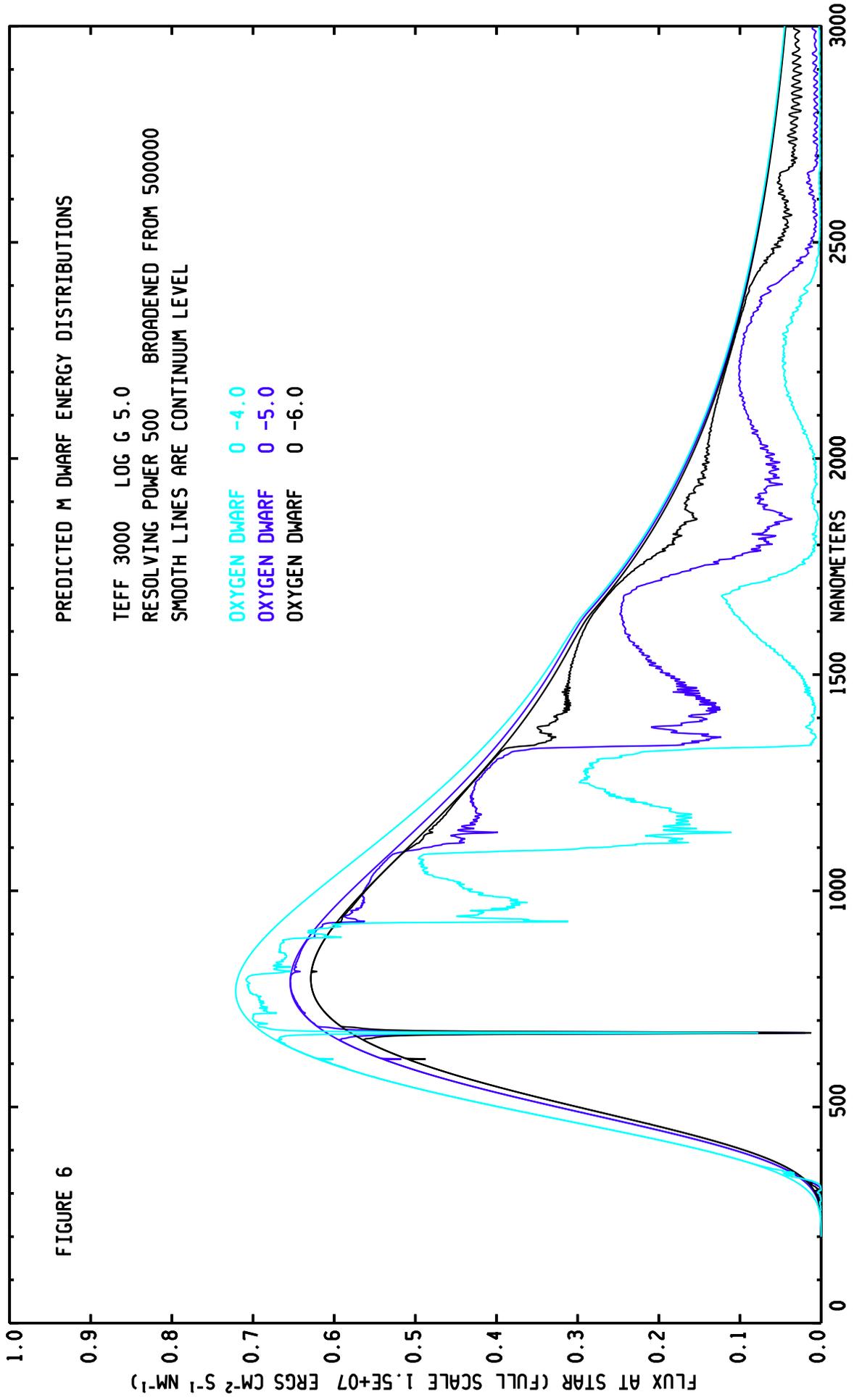
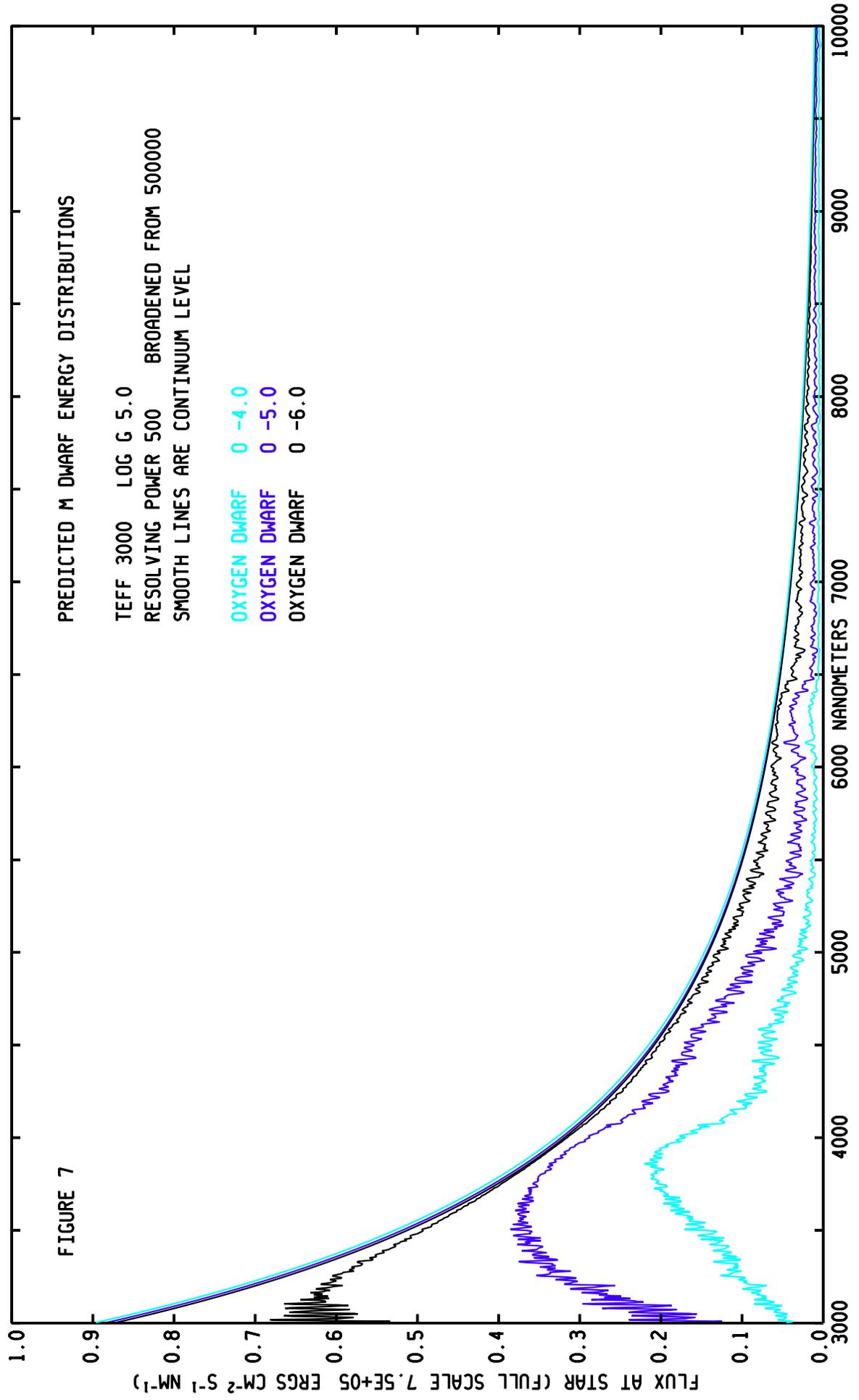
\end